# Contrastive learning-based pretraining improves representation and transferability of diabetic retinopathy classification models

**Minhaj Nur Alam, PhD[1,2,3,*], Rikiya Yamashita, MD[1], Vignav Ramesh, GED[1], Tejas Prabhune, GED[1], Jennifer I. Lim, MD[3], RVP Chan, MD[3], Joelle Hallak, PhD[3], Theodore Leng, MD[4], and Daniel Rubin, MD[1,5]**

[1]Department of Biomedical Data Science, Stanford University School of Medicine, Stanford, CA 94305, USA

[2]Department of Electrical and Computer Engineering, University of North Carolina at Charlotte, Charlotte, NC 28223, USA

[3]Department of Ophthalmology and Visual Sciences, University of Illinois at Chicago, Chicago, IL 60612, USA

[4]Department of Ophthalmology, Stanford University School of Medicine, Stanford, CA 94305, USA

[5]Department of Radiology, Stanford University School of Medicine, Stanford, CA 94305, USA

[*]Corresponding author

**Running head:** Contrastive learning for DR classification.

**Corresponding author:**

Minhaj Nur Alam, PhD

Research Fellow, Stanford University School of Medicine, 1265 Welch Road, Stanford, CA 94305

Assistant Professor, Department of Electrical and Computer Engineering, University of North Carolina at Charlotte, 9201 University City Boulevard, Charlotte, NC 28223

Phone: 708-673-5322, Email: minhajnur.alam@gmail.com

**Financial support:** This research was supported in part by seed grant from the Google Inc. (MA); Research to Prevent Blindness (TL) and NIH grant P30-EY026877 (TL).

**Conflict of Interest:** No conflicting interest for any authors.

**Word Count: 3545**

# Key points

**Question** Does self-supervised contrastive learning (CL) allow better representation learning and transferability of deep learning (DL) models to achieve higher diagnostic accuracy in diabetic retinopathy?

**Finding** In an independent clinical test set of 2,500 images, a ResNet50 model pretrained with CL weights, backed by neural style transfer augmentation had sensitivity, specificity, and area under the receiver operating characteristic (ROC) curve (AUC) of 0 0.90 (95% CI: 0.895-0.92), 0.85 (0.83-0.86), and 0.91 (0.89-0.93) respectively, for detecting referable diabetic retinopathy, compared to 0.81 (0.79-0.83), 0.74 (0.73-0.76), and 0.80 (0.78-0.82) and 0.84 (0.82-0.85), 0.79 (0.79-0.0.76), and 0.83 (0.0.80-0.85), from two state of the art baseline models (ResNet50 and InceptionV3), pretrained with Imagenet weights. At 10% labeled training data, the model retained diagnostic performance to a great extent (AUC of 0. 0.81 (0.78-0.84) in the test set, compared to 0.58 (0.56-0.64) and 0.63 (0.60-0.66) respectively, from the baseline models).

**Meaning** CL pretrained DL algorithms have high diagnostic performance for detecting referable diabetic retinopathy, even with small, labeled training datasets.

# Abstract

**Importance:** Self-supervised contrastive learning (CL) based pretraining allows development of robust and generalized deep learning (DL) models with small, labeled datasets, reducing the burden of label generation. Application of CL in diabetic retinopathy (DR) diagnosis requires further investigation.

**Objective:** To evaluate the effect of CL-based pretraining on the performance of referrable vs non referrable diabetic retinopathy (DR) classification.

**Design, Setting, and Participants:** CL is a form of self-supervision that can leverage unlabeled data to produce pretrained models. We have developed a CL-based framework with neural style transfer (NST) augmentation to produce models with better representations and initializations for the detection of DR in color fundus images. We compare our model (FundusNet – a ResNet50 architecture pretrained with NST integrated CL framework) performance with two state-of-the-art baseline models (ResNet50 and InceptionV3 architectures pretrained with Imagenet weights). We further investigate the model performance with reduced labeled training-data (down to 10%) to test the robustness of the model when trained with small, labeled datasets. The model is trained and validated on the EyePACS dataset and tested independently on clinical data from the University of Illinois,Chicago (UIC).

**Exposure:** CL pretrained DL algorithm.

**Main outcomes and measures:** The sensitivity, specificity, and area under the receiver-operating-characteristic (ROC) curve (AUC) of the algorithm (95% confidence interval, CI) for detecting referrable DR have been generated based on the reference labels created by an ophthalmologist panel.

**Results:** The validation data from EyePACS consists of 88,692 images from 44,346 individuals. After data-curation, the dataset contains 57,722 non-referrable-DR and 13,247 referrable-DR images. The independent UIC dataset contains 2500 images from 1250 patients (500 referrable and 750 non-referrable DR). Compared to baseline models, our CL-pretrained 'FundusNet' model had higher AUC(CI) values (0.91(0.898-0.930) vs 0.80(0.783-0.820) and 0.83(0.801 – 0.853) on UIC data). At 10% labeled training data, the FundusNet AUC was 0.81(0.78-0.84) vs 0.58(0.56-0.64) and 0.63(0.60-0.66) in baseline-models, when tested on the UIC dataset.

**Conclusion and relevance:** CL-based pretraining with NST significantly improves DL classification performance, helps the model generalize well (transferable from EyePACS to UIC data), and allows training with small, annotated datasets– therefore reducing ground-truth annotation burden of the clinicians.

Diabetic retinopathy (DR) is a major ocular manifestation of diabetes. According to a World Health Organization (WHO) report, it is estimated that by the year 2040, the number of diabetic patients will reach 642 million. Nearly 40-45% patients with diabetes are prone to vision impairment due DR, making the global estimate of DR patients nearly 224 million. DR can be initially asymptomatic at its non-proliferative stage (NPDR), which is characterized by the presence of micro-aneurysms. If not treated in a timely fashion, it can progress to proliferative diabetic retinopathy (PDR), leading to irreversible vision loss and blindness. The American Academy of Ophthalmology (AAO) recommends that patients with the prevalent Type II diabetes should be screened every year after the initial diagnosis [1]. However, studies show that less than 50% diabetic patients follow through and get their yearly screening, the rate being even lower (15-20%) in rural areas[2, 3]. Therefore, it is imperative to find an efficient way to improve the treatment management for DR and enable mass screening, early onset detection, and clinical diagnostics.

In recent years, researchers have successfully demonstrated machine learning (ML) and deep learning (DL) based algorithms for DR diagnosis and referrals. Especially, DL methodologies have facilitated feature extraction and DR classification with high accuracy, sensitivity, and specificity [4-16] using different imaging modalities such as fundus images, optical coherence tomography (OCT) and OCT angiography (OCTA) images. In general, such DL based DR classification pipelines require large, clean, diverse data, ground truth associated with the data, and a robust DL model (convolutional neural nets such as VGG16, ResNet, InceptionNet etc.). In case of referrable vs non-referrable DR classification, despite impressive performance showcased by these DL models, we can make two major observations: i) for wide-spread deployment, the DL models need to be more generalized, and ii) more data from different sub-populations can lead to better model training. However, this need to utilize larger, more diverse data also increases the need for ground truth generation and data labeling – which creates a large burden on clinicians. In this study, our focus is on reducing the burden of ground truth generation.

To address this goal, we have developed a self-supervised contrastive learning (CL) based pipeline for classification of referrable vs non-referrable DR. Self-supervised models like CL help a DL model learn effective representation of the data without the need for large ground truth data, the supervision is provided by the data itself. In such pipelines, a DL network is trained on a primary task (for representation learning, which requires no ground truth), and then the weights from that task are transferred to a secondary target task (i.e., classification, which requires smaller set of data and ground truth). For example, if a model is trained to solve an image puzzle, it does not need ground truth (the input image is the reference ground truth). But by learning to solve the puzzle, the model learns effective representation and the characteristic features from the image. This model's weight can be then used for image classification task – yielding high classification performance with smaller data and ground truth [17]. In this paper, we present 'FundusNet' – a CL based framework that achieves high classification performance even with smaller sets of fundus image data. In addition to that, we introduce a neural style transfer (NST) - based image augmentation technique, that effectively improves the representation learning capability of the CL network from fundus images. The model with FundusNet weights is independently evaluated on external clinical data, which achieves high sensitivity and specificity, when compared to a baseline model. The CL model also performed well even when the labelled dataset was reduced to 10% of its original size, suggesting the potential of CL to train models for DR diagnosis using small, labeled datasets.

## Methods

### Study design and participants

This study was approved by the institutional review board of Stanford University and the University of Illinois at Chicago (UIC) and was in compliance with the ethical standards stated in the declaration of Helsinki. This multi-center cross-sectional study was primarily conducted at Stanford University School of Medicine. The testing data from UIC was shared in encrypted cloud drive with researchers at Stanford. For training and developing the CL based pretraining and referrable vs non-referrable DR classifier, we used the EyePACS dataset from Kaggle (88,702 fundus photographs, EyePACS, California). The final DR classifier model was tested on an independent dataset from UIC. The training data from EyePACS contained retinal fundus photographs from patients with varying degrees of severity of DR. The dataset had 71,548 non-referable (65,343 no DR, 6205 mild NPDR) and 17,154 referable (13,153 moderate NPDR, 2087 severe NPDR, and 1914 PDR) DR images with varying resolutions from 433 x 289 up to 5184 x 3456 pixels. Images from the dataset are already labeled with stages of DR (0: no DR, 1: mild, 2: moderate, 3: severe non-proliferative DR (NPDR), and 4: proliferative DR (PDR)), following the diagnostic criteria for DR. For our project, we define referable DR as data which have labels of moderate NPDR and above (label >= 2). We excluded images that had motion artifacts, were too dark or blurry to confidently stage disease, and images with missing fovea or optic disk. An automated image quality assessment algorithm was used to identify the images that fit the exclusion criteria [18, 19]. The testing dataset from UIC contained 2500 fundus photographs from patients with DR, recruited from the UIC retina clinic (1000 referable and 1500 non-referable DR). This was retrospective data of type II diabetes patients who underwent retinal imaging at the clinic. The patients are thus representative of a university population of diabetic patients who require imaging for management of diabetic macular edema and DR. Images of both eyes were taken. Subjects with macular edema, previous history of eye diseases, and vitreous surgery were excluded from the study. The patients were classified by severity of DR according to the Early Treatment Diabetic Retinopathy Study staging system, which was then converted to class labels of no DR, mild, moderate, and severe NPDR, and PDR. The grading was done by retina specialist on dilated patients who were examined using a slit-lamp fundus lens, and technicians did not contribute to the grading of the patients. All patients in this study provided written informed consent and did not receive any compensation or incentives to participate.

## Framework for contrastive learning-based pretraining

Our FundusNet framework consists of two primary steps. First, we perform self-supervised pretraining on unlabeled fundus images from the training dataset using contrastive learning to learn visual representations. Once the model has been trained, the weights are transferred to a secondary classifier model for supervised fine-tuning on labeled fundus images. Figure 1 describes a summary of the framework.

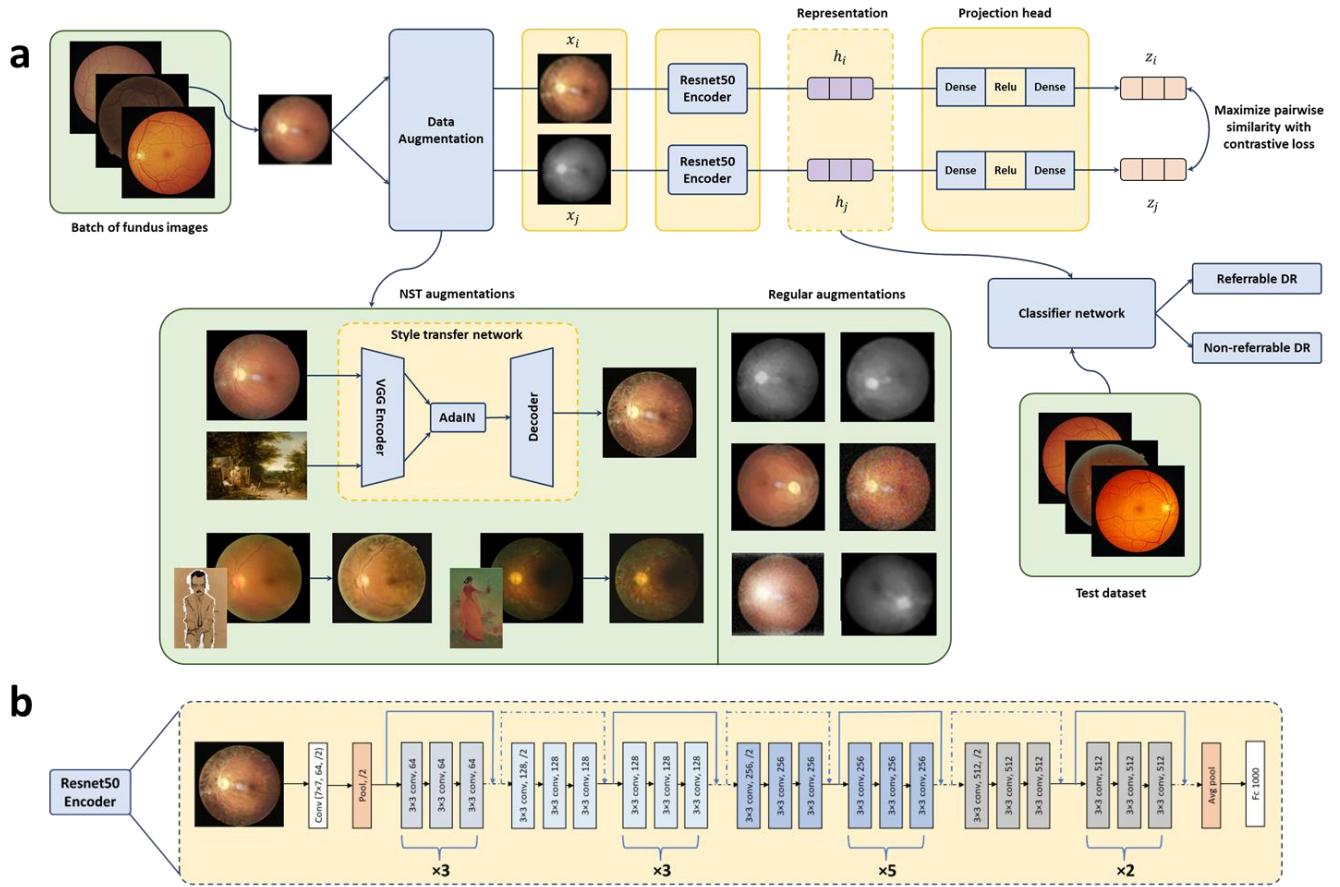

Figure 1: (a) A framework for contrastive learning based pretraining for referrable vs non-referrable diabetic retinopathy classification. NST denotes neural style transfer. The training utilizes the EyePACS dataset, whereas the test dataset comes from the UIC retinal clinic. The representations $h_i$ and $h_j$ are used as transfer learning weights for the classifier network after the contrastive learning pipeline is optimized, i.e., the contrastive loss has reached its minimum value. The AdaIN refers to adaptive instance normalization that allows real time style transfer, described in [20, 21]. Regular augmentation refers to augmentations used in the original SimCLR paper [22] (flipping, rotation, color distortion); (b) Architecture of the ResNet50 encoder.

To teach our model visual representations effectively, we adopt a SimCLR framework [22], which is a recently proposed self-supervised approach that relies on contrastive learning. In this method, the model learns representations by maximizing the agreement between two differently augmented versions of the same data using a contrastive loss (more details on contrastive loss is provided in supplemental material).

This contrastive learning framework (Figure 1a) attempts to teach the model to distinguish between similar and dissimilar images. Given a random sample of fundus images, the FundusNet framework takes in each image X, augments them twice, creating two versions of the input image $X_i$ and $X_j$. The two images are encoded via a ResNet50 network (Figure 1b), generating two encoded representations $h_i$ and $h_j$. These two representations are then transformed via a non-linear multi-layer perceptron (MLP) projection head, yielding two final representations, $z_i$ and $z_i$, which are used to calculate the contrastive loss. Based on the loss on each augmented pairs generated from a batch of input images, the encoder and projection head representations improve over time and the representations obtained place similar images closer in the representation space. The CL framework contains a Resnet50 encoder (containing convolutional neural network and pooling layers with skip connections) with a projection head (dense and Relu layers) that maps the representation. The batch size of the CL pretraining pipeline has been demonstrated to have significant effect on the model pretraining [22-24] and therefore, the performance of the target model. To test this, we trained our FundusNet model for bath sizes 32 through 4096 (step size 32). The model is trained for 100 epochs or until the loss function saturates.

## Improving representation learning through neural style transfer (NST)

One of the key findings from CL based self-supervised pretraining is that augmentation and transformation are key to better representation learning. As we adopted and modified the SimCLR framework, which was originally used on natural images, we found that regular image augmentation techniques such as flipping, rotating etc. did not generate a good representation ($Z_i$ and $Z_i$ in figure 1a) from fundus images. A study by Geirhos et.al. [21] demonstrated that CNNs used in computer vision tasks are often biased towards texture, compared to global shape features that are primarily used by humans for distinguishing classes. Increasing shape bias by randomizing texture environments can be a useful way to improve accuracy and generalizability of a CNN model. NST manipulates the low-level texture representation of an image (style) but preserves the semantic content. NST has been previously demonstrated to improve robustness to domain shift in CNNs for computer vision tasks [25, 26]. In our study, we integrated an NST-based augmentation technique into the CL pipeline, based on convolutional style transfer from non-medical style sources (i.e., art, painting etc.). The NST replaces the style of the fundus images (primarily texture, color and contrast) with the randomly selected non-medical images. However, it preserves the semantic contents (global objects, shapes like microaneurysm, vasculature etc.) of the image required for better disease detection. The NST convolution methodology was adopted from AdaIn style transfer [20, 21]. The style source was artistic paintings from Kaggle's 'Painter by Numbers' dataset (79,433 paintings), downloaded via https://www.kaggle.com/c/painter-by-numbers. In the CL pretraining, the NST based augmentation was combined with the regular augmentation techniques such as rotation, flipping, color distortion, crops with resize, and gaussian blur. A higher probability (70%) of augmentation through NST was defined in the pretraining protocol. To compare the performance improvement of detecting referrable DR due to integration of NST into our pipeline, we also trained a CL framework with just the original SimCLR augmentations [22].

**Referrable vs non-referrable DR classification**

Using the weights of the pretrained network as initializations, we trained an end-to-end supervised model for a downstream DR classification task (referrable vs non-referrable DR). We trained a ResNet50 encoder network with standard cross-entropy loss, a batch size of 256, ADAM optimizer and random augmentations (gaussian blurring, resizing, rotations, flipping, and color distortions). The fundus images were resized to 224 x 224 pixels during this stage. This pixel dimension was optimized based on the tradeoff between image resolution and memory limitation during model training. To compare our FundusNet results, we also trained two separate fully supervised baseline models (ResNet50 and InceptionV3 encoder networks, both initiated with Imagenet weights). Both the baseline models are based on based on DL models in literature that have achieved state of the art diagnostic accuracy in detecting referrable DR [16, 27].  Standard hyperparameter search (learning rate (logarithmic grid search between $10^{-6}$ and $10^{-2}$), optimizer (ADAM, SGD), batch size (32, 64, 128, 256)) and training protocols were maintained for the FundusNet and both baseline networks.

To further investigate whether the CL pretrained model performs well with smaller training data (and ground truth), we reduced the training dataset gradually from 100% to 10% (10% step size) and conducted the downstream classification training for both the CL and Imagenet pretrained baseline models. After identifying the best hyperparameters and fine tuning the models for each experiment, we chose the model that had the best performance on validation dataset (5-fold cross validation). The final optimal models were tested on an independent testing dataset from UIC. In terms of encoder networks, we compared three types of encoder networks in our experiment (VGG, ResNet, and Inception architectures).

**Statistical Analysis**

Our primary metrics to evaluate the model performance was the area under the curve (AUC) with 95% confidence intervals (CI). For each experiment, sensitivity and specificity of the CNN classifiers were also computed across probability thresholds to plot the receiver operating characteristic (ROC) curves and calculate AUC. For individual AUC, statistical comparisons were performed using DeLong's test[28]. We also compared the age, sex and hypertension distribution among different DR cohorts using one-way, multi-label analysis of variance (ANOVA) test.

# Results

This study used EyePACS dataset for the CL based pretraining and training the referrable vs non-referrable DR classifier. EyePACS is a public domain fundus dataset which contains 88,692 images from 44,346 individuals (both eyes, OD and OS), and the dataset is designed to be balanced across races and sex. After removing the fundus images that met the exclusion criteria, the final training dataset contains 70,969 fundus images, of which 57,722 are non-referrable DR and 13,247 are referrable DR. An independent testing dataset from UIC retina clinic is used for the target task of DR classification. This dataset contains 2500 images from 1250 patients (both eyes OD and OS). Among 1250 subjects (mean [SD] age, 53.37 [11.03]), 818 were male (65.44%) and 432 were female (34.56%). The detailed demographic information of the subjects from UIC is in table 1. There was no statistically significant difference in the distribution of age, sex, and hypertension between non-referrable and referrable DR groups (ANOVA, P = 0.32, 0.18, 0.59 respectively).

Table 1. Demographics characteristics of non-referrable and referrable DR subjects from the testing dataset at UIC.

|  | Non-referrable DR | Referrable DR |
|---|---|---|
| Number of subjects | 750 | 500 |
| Sex (male/female) | 458/292 | 360/140 |
| Age (mean ± SD) | 50.36 ± 10.64 | 56.37± 11.84 |
| Age range | 28-73 | 32-78 |
| Duration of disease (years) | 14.23 ± 10.22 | 19.32 ± 12.94 |
| Diabetes type | Type II | Type II |
| Insulin dependent(Y/N) | 117/633 | 389/111 |
| HbA1C, % | 6.6 ± 4.1 | 8.1 ± 3.2 |
| HTN prevalence, % | 69 | 81 |

DR: diabetic retinopathy, SD: standard deviation, HbA1C: Glycated hemoglobin, HTN: hypertension

The FundusNet model pretrained with CL and style transfer augmentation achieved an average AUC of 0.91 on the independent test dataset from UIC, outperforming the state-of-the-art baseline models (ResNet50 and InceptionV3) trained with Imagenet weights [16] (AUCs of 0.80 and 0.83, respectively) (Table 2). The significant performance difference in the testing test compared to the baseline model indicates that the FundusNet model generalized better through our pretraining framework. The NST augmentation further allowed learning of more discriminative visual representations of retinal pathologies, improving the overall classification performance (0.91 (95% CI: 0.898-0.93) with NST augmentation vs 0.83 (95% CI: 0.80-0.85) with original SimCLR augmentations [22]).

Table 2: Classification performance of FundusNet framework for referrable vs non-referable DR.

| Models | AUC (95% CI) | P-value (vs FundusNet) | Sensitivity (95% CI) | P-value (vs FundusNet) | Specificity (95% CI) | P-value (vs FundusNet) |
|---|---|---|---|---|---|---|
| FundusNet | 0.91 (0.898 – 0.930) | Ref | 0.90 (0.895 – 0.917) | Ref | 0.85 (0.830 – 0.862) | Ref |
| Baseline1 (ResNet50) | 0.80 (0.783 – 0.820) | P < 0.001 | 0.81 (0.793 – 0.834) | P < 0.001 | 0.74 (0.731 – 0.758) | P < 0.005 |
| Baseline2 (Inception V3) | 0.83 (0.801 – 0.853) | P < 0.001 | 0.84 (0.822 – 0.848) | P < 0.001 | 0.79(0.786 – 0.819) | P < 0.05 |

DR: diabetic retinopathy, CI: confidence interval; AUC: area under the ROC curve; Ref: reference; P value from measuring statistical significance using DeLong's test for comparing pairwise AUCs.

To investigate the label-efficiency of the FundusNet model, we trained our model on different fractions of the labeled training data and tested each resulting model on the test dataset. We compared this to the performance of the baseline models. Fine-tuning experiments were conducted on five-folds training data and the results were averaged. Figure 2 shows how the performance varies using the different label fractions for both the FundusNet and baseline models. Figure 2 shows the performance comparison on testing dataset. We observe that the CL pretrained FundusNet model retains AUC performance even when the labels are reduced up to 10%, whereas there is significantly smaller performance of the baseline models. When reducing the amount of training data from 100% to 10% of the data, the AUC for FundusNet drops from 0.91 to 0.81 when tested on UIC data, whereas the drop is larger for the baseline models (0 0.80 to 0.58 for the ResNet50 and 0.81 to 0.63 for the InceptionV3 model). Importantly, the FundusNet model is able to match the performance of the baseline models using only 10% labeled data when tested on independent test data from UIC (FundusNet AUC 0.81 when trained with 10% labelled data vs 0.80 and 0.81, respectively, for baseline models trained with 100% labelled data).

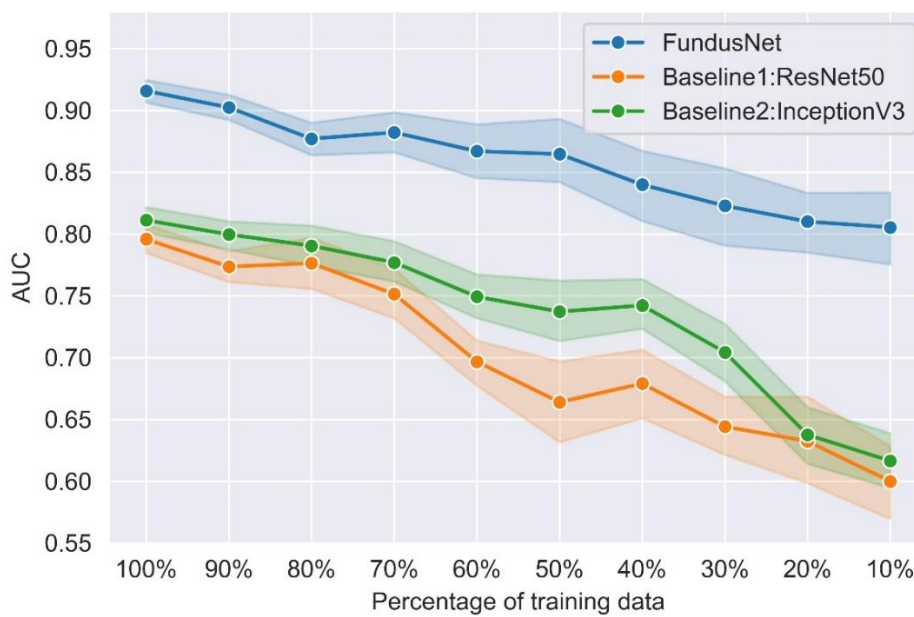

Figure 2: AUC for referrable vs non-referrable DR classification tested on independent test data from UIC for FundusNet vs two baseline models with varied percentages of training data.

Table 3: Effect of batch size on FundusNet performance on detecting referrable DR.

| Batch size | AUC (SD) – test dataset | P value (vs 32 batch size) | P value (vs 2048 batch size) |
|---|---|---|---|
| **32** | 0.77 (0.039) | REF | <0.001* |
| **64** | 0.78 (0.058) | 0.18 | <0.001* |
| **128** | 0.78 (0.045) | 0.16 | <0.001* |
| **256** | 0.80 (0.077) | 0.06 | <0.001* |
| **512** | 0.86 (0.082) | 0.02* | <0.001* |
| **1024** | 0.87 (0.034) | < 0.01* | <0.01* |
| **2048** | 0.91 (0.014) | < 0.001* | REF |
| **4096** | 0.91 (0.011) | < 0.001* | 0.08 |

AUC: area under the ROC curve; SD standard deviation; Ref: reference; '*' indicates significant difference; P value from measuring statistical significance using DeLong's test for comparing pairwise AUC values among the batch size (all vs batch size 32 in third column, and all vs batch size 2048 in fourth column). This pairwise comparison was done to show whether the AUCs coming from experiments with all the batch sizes are significantly different that AUCs coming from experiments with batch size 32 (initial batch size) and 2048 (optimal batch size).

In the experiment to evaluate the optimal batch size for CL pretraining, we observed that CL frameworks learned better image representations when there were higher number of negative examples in a batch (i.e., augmented image pairs generated from other images in a batch), therefore, higher batch size yielded better performance (Table 3; AUC of 0.77 for batch size 32 vs 0.91 for batch size 2048 on test dataset). However, high batch size also means the need for larger compute resources. We observed that at batch size 4096, the AUC did not improve significantly, so the optimum batch size was chosen as 2048. In terms of encoder networks, compared to VGG and Inception architectures, ResNet50 provided the best classification performance in both validation and test dataset (Supplemental table 1).

# Discussion

In this paper, we present a self-supervised CL based pipeline, FundusNet, for improving the performance, of referrable vs non-referrable DR classification over previously published baseline models [16, 27]. CL improves representation and transferability of DR classification models. Our key contributions are: i) the CL based pretraining method significantly outperformed the baseline models pretrained with Imagenet weights; ii) we proposed a novel NST based augmentation within the CL framework that enhances the capability of the model to learn fundus data representation for robust classification performance; and iii) the CL pretrained models performed well even when we reduced the amount of training data by 90%. The FundusNet network with CL pretraining shows potential for integrating self-supervision in deep learning frameworks for developing robust diagnostic models with reduced amount of data and associated ground truth. Previous models for the diagnosis of DR used Imagenet weights to initiate their models [10, 11, 27, 29-31]. While transfer learning through Imagenet can be useful, recent work has shown that the role of Imagenet weights for medical imaging tasks can be limited [32, 33] since it was developed and validated on natural images, and pretraining on in-domain medical image data (i.e., FundusNet weights) can be more effective [33] compared to Imagenet.

The FundusNet model achieves high sensitivity and specificity in referrable vs non-referrable DR classification (Table 2) and performed significantly better than the baseline models (ResNet50 and InceptionV3 [16, 27]) on the independent test dataset (AUC of 0.91 vs 0.80 and 0.81 respectively) and on EyePACS 5-fold validation data (AUC of 0.95 vs 0.92 and 0.94 respectively), suggesting that the CL model generalized better. Several previous studies have demonstrated great success to detect referrable DR using deep learning approaches. Gulshan et al. developed a model with 128,175 images and validated the model on 9963 fundus images high level of performance for classifying referable DR (AUC = 0.99) [7]. Ting et al. trained their deep learning model using 73,370 images and reported excellent results for referable vs non referrable DR (AUC of 0.936) [14]. Li et al. trained their deep learning system on 71,043 images and validated them in a real-world dataset of 35,201 images (AUC of 0.955 for vision-threatening

DR) [27]. To achieve such excellent diagnostic accuracy, these works used a large diverse training dataset (ranging from 70,000 to 200,000 images) with class labels created by multiple clinicians. Our CL framework could potentially reduce the need for large training datasets. On the independent test data from UIC, the FundusNet model matched the AUC of the two baseline models with only 10% of labeled training data (Figure 2).

In our study, we not only adapted contrastive learning into our training pipeline, we also further modified the pipeline with a novel NST based augmentation. NST is an image style transformation algorithm that manipulates the low-level texture representation of an image while preserving the semantic content. NST has been previously demonstrated to improve robustness to domain shift in CNNs for computer vision tasks [20], and we have adopted it into our FundusNet framework. This is a unique application of style transfer from medically irrelevant artistic images in ophthalmic diagnostics. The NST augmentation facilitated learning discriminative visual representations of retinal pathologies by encouraging the model to learn global shape based structural characteristics of vasculature and other pathologies in the retina. NST works well as an augmentation method because the style transfer can introduce in a wide variety of textures when training classifiers using medical images (which often have more uniform color and texture distribution), that can improve classifier performance. Therefore, compared to conventional augmentations (such as the ones used in the original SimCLR [22]), the model generalizes better when augmented with NST using texture-rich artistic images (AUC of 0.91 vs 0.83 when FundusNet used NST vs did not used NST on top of regular augmentations).

Our study has several limitations. First, we did not optimize the NST hyperparameters. The use of NST as an augmentation technique is new and requires further investigation to optimize the style coefficients. Another limitation of our approach is that a large batch size is required for training of the CL model. Self-supervised frameworks like SimCLR and MoCo reported the need for larger batch size [22-24] because CL training requires a large number of negative samples in a batch to calculate contrastive loss efficiently. This can impair computational efficiency. Furthermore, our model was tested on only one external clinical site at UIC, which may not be sufficiently representative of a broader population. Further validation of the developed FundusNet framework is required in future studies.

In conclusion, we have introduced a self-supervised CL based framework for referrable vs non-referrable DR classification that includes NST augmentation for learning effective pathological and visual representations from retinal fundus images. Our experiments demonstrate that our CL based pretraining yields significant improvements of DR classification compared with the baseline models in independent testing data. Importantly, our CL model is able to be trained using only 10% of the training data that could reduce the annotation burden for clinicians in producing training datasets. Ultimately, our CL method could be useful for developing classification models for clinical use when validated sufficiently.

**Code and data availability:**

The code is open-source. All the code and FundusNet weights are available in
https://github.com/malam8/FundusNet-DR. The validation dataset from EyePACS is available at
https://www.kaggle.com/c/diabetic-retinopathy-detection/data. The test dataset from UIC can be made available upon request.

**Contributors:**

This project was led by MA, who supervised and developed the contrastive learning framework and DR detection models. VR and TP helped with the contrastive learning training; RY helped with the neural style transfer integration into the contrastive learning pipeline. The data from UIC came from patients of JIL and RVPC; JH, JIL, and RVPC helped with data curation, collection, and annotation to referrable vs non-referrable diabetic retinopathy. MA, TL, and DR at Stanford contributed to the conception and design of this study. All the authors helped with interpreting the results, designing experiments, and writing of this manuscript. All the authors read and approved the final draft of the manuscript.

**Declaration of Interests:**

The authors declare no relevant financial interests to this project.